\RequirePackage{fix-cm}
\documentclass[smallextended]{svjour3}       % onecolumn (second format)
\smartqed  % flush right qed marks, e.g. at end of proof
\usepackage{graphicx}
\usepackage{epsfig}
\usepackage{amsmath}
\usepackage{color}
\usepackage{array}

 \journalname{Bulletin of Mathematical Biology}
\usepackage{natbib}
\bibpunct{(}{)}{;}{a}{}{,} 

\begin{document}
%\title{Mathematical Foundations of Anticoagulant Adjuvant Therapies in Glioblastoma Multiforme}
\title{Hypoxic Cell Waves around Necrotic Cores in Glioblastoma: A Biomathematical Model and its Therapeutic Implications}

\thanks{We wish to thank Marcial Garc\'{i}a Rojo and Cristina Murillo (Servicio de Anatom\'{\i}a Patol\'ogica, Hospital General Universitario de Ciudad Real, Spain) for fruitful discussions. We wish to thank D. Brat (Emory University, USA) for his support and for giving us the photographs used for Figs. \ref{fig:minipalisade_Brat} and \ref{fig:largepalisade_Brat}. 
This work has been supported by grants MTM2009-13832 (Ministerio de Ciencia e Innovaci\'on, Spain) and PEII11-0178 (Junta de Comunidades de Castilla-La Mancha, Spain). }

%\subtitle{Do you have a subtitle?\\ If so, write it here}

%\titlerunning{Modelling Perinecrotic Palisades in Glioblastomas} 
\titlerunning{Cell Waves around Necrotic Cores in GBM} 

\author{ Alicia Mart\'{\i}nez-Gonz\'alez \and Gabriel F. Calvo \and Luis A. P\'erez Romasanta \and V\'{\i}ctor M. P\'erez-Garc\'{\i}a %etc.
}

\authorrunning{Alicia Mart\'{\i}nez-Gonz\'alez \emph{et al.}} % if too long for running head

\institute{A. Mart\'{\i}nez-Gonz\'alez, V. M. P\'erez-Garc\'{\i}a \at
              Departamento de Matem\'aticas, E. T. S. I. Industriales and Instituto de Matem\'atica Aplicada a la Ciencia y la Ingenier\'{\i}a, Universidad de Castilla-La Mancha, 13071 Ciudad Real, Spain. \\
              \email{alicia.martinez@uclm.es, victor.perezgarcia@uclm.es}           %  \\
%             \emph{Present address:} of F. Author  %  if needed
\and
           G. F. Calvo \at
                          Departamento de
Matem\'aticas, E. T. S. I. Caminos, Canales y Puertos and Instituto de Matem\'atica Aplicada a la Ciencia y la Ingenier\'{\i}a, Universidad de Castilla-La
Mancha, 13071 Ciudad Real, Spain.
\email{gabriel.fernandez@uclm.es}
 \and
           L. A. P\'erez Romasanta \at
Servicio de Oncolog\'{\i}a Radioter\'{a}pica. Hospital General Universitario de Ciudad Real, 13005 Ciudad Real, Spain.
\email{luisp@sescam.jccm.es}
}

\date{Received: date / Accepted: date}
% The correct dates will be entered by the editor

\maketitle

\begin{abstract}
Glioblastoma is a rapidly evolving high-grade astrocytoma that is distinguished pathologically from lower grade gliomas by the presence of necrosis and microvascular hiperplasia. Necrotic areas are typically surrounded by hypercellular regions known as ÒpseudopalisadesÓ originated by local tumor vessel occlusions that induce collective cellular migration events. This leads to the formation of waves of tumor cells actively migrating away from central hypoxia. We present a mathematical model that incorporates the interplay among two tumor cell phenotypes, a necrotic core and the oxygen distribution. 
Our simulations reveal the formation of a traveling wave of tumor cells that reproduces the observed histologic patterns of pseudopalisades.
Additional simulations of the model equations show that preventing the collapse of tumor microvessels leads to slower glioma invasion, a fact that might be exploited for therapeutic purposes.

%\PACS{}%87.19.xj \and 87.10.Ed \and 87.17.Aa} 
\end{abstract}

\section{Introduction}
\label{intro}

\begin{figure}[t]
\begin{center}
%\vspace*{-2mm}
%\hspace*{1mm}
\includegraphics[scale=0.25]{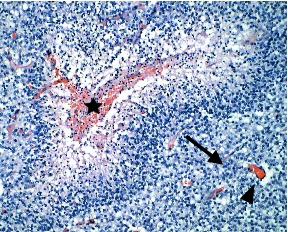} 
\end{center}
\caption{{\bf Pseudopalisading structures in Glioblastoma Multiforme.} This histopathologic sample shows the typical features of pseudopalisades in GBM. Hypercellular regions appear around  centrally degenerating vessels (\textit{star}). The standard interpretation \citep{pseudopalisading1} is that cells migrate away from the degenerated vessels (hypoxic) areas and move towards functional vascular supplies such as the one marked with an arrow head in the photograph. 
\label{fig:minipalisade_Brat}}
\end{figure}

Malignant gliomas are the most common and lethal type of primary brain tumor. Survival for patients with glioblastoma multiforme (GBM), the most aggressive and prevalent  WHO grade IV astrocytic glioma \citep{WHO,Wen}, although individually variable, is about 12 months after diagnosis, using the standard of care which includes surgery to resect as much tumoral tissue as possible, radiotherapy and chemotherapy (temozolamide) \citep{Huse,Meir,Clarke,Reardon,Pruitt}. GBM is a rapidly evolving astrocytoma that is distinguished pathologically from lower grade gliomas by the presence of necrosis in the central regions of the tumor and microvascular hyperplasia \citep{WHO}.  Another distinctive signature of GBM is the fact that necrotic foci are frequently surrounded by hypercellular regions known as {\em pseudopalisades}. An example of these hypercellular perinecrotic structures is shown in Figure \ref{fig:minipalisade_Brat}. 

\par

The fact that GBM is one of the most highly vascularized human tumors seems to contradict that its microcirculation is functionally very inefficient \citep{Jensen}. However, it is well known that the tumor microenvironment is remarkably different from that of normal tissue; its chaotic vasculature leads both to a decrease in the supply of oxygen and essential nutrients as well as in the removal of waste products. 

Tumor hypoxia (deficit in oxygenation) is a feature encountered in most solid tumors \citep{Semenza03,Vaupel,Bristow,Dewhirst}, albeit with an incidence and severity that varies among patients. It is generally recognized as a negative clinical prognostic and predictive factor owing to its involvement in various cancer hallmarks such as resistance to cell death, angiogenesis, invasiveness, metastasis, altered metabolism and genomic instability \citep{Hanahan2011,Wilson}. Hypoxia displays a central role in tumor progression and resistance to therapy (chemo- and radioresistance), specially in GBM \citep{Jensen,Pope}. Low oxygen levels are observed both in the near vicinity of the tumor vasculature, as acute or cycling hypoxia \citep{Bristow,Dewhirst}, or at longer distances of about 150 $\mu$m from the feeding blood vessels giving rise to areas of chronic or diffusion-limited hypoxia \citep{Bristow}. Angiogenesis emerges then in response to the imbalance of proangiogenic growth factors that are released by hypoxic cells in the tumor such as vascular endothelial growth factor (VEGF) relative to anti-angiogenic growth factors (e.g., angiostatin) present in the tumor microenvironment \citep{Carmeliet,Ebos}. The end result of VEGF signaling in tumors is the production of immature, highly permeable blood vessels with subsequent poor maintenance of the blood brain barrier and parenchymal edema \citep{Jain2007}. Thus, microvascular hyperplasia may coexist with moderate or high levels of hypoxia. Actually, for GBM it has been shown that microvascular hyperplasia is spatially and temporally associated with pseudopalisading necrosis and is believed to be driven by hypoxia-induced expression of proangiogenic cytokines such as VEGF \citep{pseudopalisading2}.

\par

In general, GBMs are pathologically heterogeneous with central regions of necrosis surrounded by pseudopalisading cells and hypervascularized regions under moderate levels of hypoxic stress \citep{Gorin}. In addition, the periphery of the GBM typically shows tumor cells infiltrating into the normal brain. This invasion often extends to the adjacent cortex and through the corpus callosum into the contralateral hemisphere, being a major reason for treatment failure: migrating cells are not eliminated by surgical resection and cause tumor recurrence \citep{Onishi}. Thus, high grade astrocytic tumors typically recur in less than six months after surgery \citep{Giese}. 
\par

Emerging evidence from clinical observations suggests that the origin of pseudopalisade formation is a vaso-occlusive event where the local blood vessels no longer provide the necessary oxygen supply to the tumor microenvironment, resulting in a population of migrating hypoxic cells responsible for the formation of the pseudopalisades (see Figure \ref{fig:minipalisade_Brat}). Establishing a clear mechanism underlying pseudopalisade formation and its role in necrosis is critical since their formation  could potentially precede all forms of coagulative necrosis in GBM \citep{vaso-occlusive,pseudopalisading1}, responsible for the broad necrotic areas detected, for example, by standard magnetic resonance imaging of brain tumors. In fact, pathologic examinations indicate that hypoxia and necrosis development within high grade astrocytomas could arise secondary to vaso-occlusion. Pseudopalisades would then represent a wave of tumor cells migrating away from central hypoxia \citep{pseudopalisading2}.
\par

Experimental studies have hypothesized that pseudopalisade formation could provide a mechanism accounting for the rapid clinical progression of GBM,
acting as a link among the underlying vascular damage, the development of hypoxia and hypoxia-induced angiogenesis, leading to necrosis and accelerated outward tumor expansion in the form of a migrating wave of cells. Previous works have proposed that pseudopalisade formation would be the result of a multistep process \citep{pseudopalisading1,pseudopalisading2,vaso-occlusive}: First, in anaplasic astrocytoma (WHO grade III) or {\em de novo} GBM in its initial stages, tumor cells proliferate and infiltrate through the parenchyma and receive oxygen and nutrients via the intact native blood vessels. Secondly, vascular insult occurs as a result of uncontrolled tumor growth causing endothelial injury and vascular leakiness. Both endothelial injury and the expression of procoagulant factors by the neoplasm result in intravascular thrombosis and increasing hypoxia in the regions surrounding the vessel \citep{Rahman}. Subsequently, tumor cells begin to migrate away from hypoxia, creating a peripherally moving wave that is seen microscopically as pseudopalisading cells. This leads to an expansion of the zone of hypoxia and central necrosis, whereas the hypoxic tumor cells in pseudopalisades secrete proangiogenic factors (e.g. cytokines VEGF and IL-8). 

\par

\begin{figure}[t]
\begin{center}
%\vspace*{-2mm}
%\hspace*{1mm}
\includegraphics[scale=0.45]{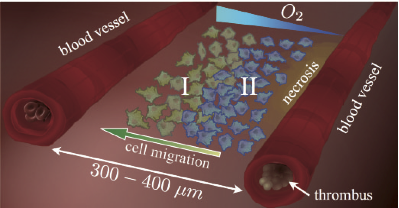} 
\end{center}
\caption{{\bf Glioblastoma microenvironment modeled in this work.} The system modeled consists of two blood vessels and an evolving embedded population of tumor cells whose phenotype changes according to the levels of oxygenation. Upon thrombosis of one of the blood vessels (the right one), due to excessive tumor cell growth and the secretion of thrombotic factors, decreasing levels of local oxygenation arise that promote a massive migration of hypoxic glioma cells (labeled as II) towards better oxygenated areas. In the process, the hypoxic glioma cells infiltrate among the normoxic glioma cells (labeled as I), which are closer to a functional vessel (the left one), creating a transient hypercellularity region (a pseudopalisade). As the pseudopalisading front of glioma cells enlarges around the thrombosed vessel, perivascular necrosis becomes more prominent.\label{fig:scheme}}
\end{figure}

In this work we study the formation of hyper cellular regions in perinecrotic areas in high grade gliomas. We put forward a mathematical model that incorporates the spatio-temporal interplay among two tumor cell phenotypes corresponding to well oxygenated (normoxic) and hypoxic cells, a necrotic core and the oxygen distribution. Our physio-pathologic scenario is depicted in Figure \ref{fig:scheme} and consists of a tumor cell population embedded within two blood vessels arranged in a flat domain. Our numerical simulations reveal the formation of a superimposed traveling wave of hypoxic cells that qualitatively reproduces the experimentally observed patterns and may be used to obtain some hints for the timescales of palisade formation, lifetime and persistence among other prognostic metrics. We also explore the dynamics of tumor spreading under delayed vascular injury and show that somehow, contrary to naive intuition, in the framework of our model preventing vessels from breaking leads to slower tumor invasion speeds that might be explored for therapeutic purposes.
\par
%%%%%%%%%%%%%%
\section{The Model}
\label{sec_model}

\subsection{Glioblastoma compartment dynamics}  

Our modeled system comprises three different compartments; two different coupled tumor cell subpopulations corresponding to the two dominant phenotypes well described in GBM \citep{DeBerardinis,Giese,Keunen,Onishi} and a necrotic compartment. Of the two populations, the first one, $C_n$, consists of normoxic, proliferative, less mobile cells (typically located close to functional blood vessels), whereas the second one, $C_h$, is composed of hypoxic, less proliferative and more mobile cells. Both phenotypes compete for space and resources (in our case oxygen). The cell loss is included in a third compartment, $C_d$, of necrotic tissue. The equations governing the interplay among these three densities are as follows
\begin{subequations}
\label{themodel}
\begin{eqnarray}
	\frac{\partial C_n}{\partial t} & = & D_n \nabla^2 C_n + \frac{1}{\tau_n} \left(1 - \frac{C_n + C_h+C_d}{C^{\left(M\right)}}\right)C_n + \frac{1}{\tau_{hn}} S_{hn}C_h  \nonumber \\ & & -  \frac{1}{\tau_{nh}} S_{nh}C_n, \label{normox} \\
	\frac{\partial C_h}{\partial t} & = & D_h \nabla^2 C_h +\frac{1}{\tau_h} \left(1 - \frac{C_n + C_h+C_d}{C^{\left(M\right)}}\right)C_h  -  \frac{1}{\tau_{hn}} S_{hn}C_h   \nonumber \\ & & + \frac{1}{\tau_{nh}} S_{nh}C_n-   \frac{1}{\tau_{hd}} S_{hd} C_h, \label{eqhipox}\\	
	\frac{\partial C_d}{\partial t} & = & \frac{1}{\tau_{hd}} S_{hd} C_h. \label{death} 
\end{eqnarray}
\end{subequations}
The first terms in Eqs. (\ref{normox}) and (\ref{eqhipox}) account for the cellular motility. Migration in gliomas is not simple and in fact  
many have proposed that the highly infiltrative nature of human gliomas recapitulates the migratory behavior of glial progenitors \citep{Suzuki,Dirks}. Here we assume, as in most models \citep{V1,V2,V3,V4,V5,Swanson,V6}, that glioma cell invasion throughout the brain is basically governed by a standard Fickian diffusion process. 
\par
It is believed that glioma cells -as many other types of cells- do not move and proliferate simultaneously \citep{Giese}. This is referred to as the migration/proliferation dichotomy and leads to the fact that highly motile cells should exhibit low proliferation rates. The switch between proliferative and invasive phenotype cannot be only mutation driven \citep{Hatzikirou,Onishi} and it has been suggested  that invasive glioma cells are able to revert to a proliferative cellular program and vice versa, depending on the environmental stimuli \citep{Giese,Keunen}. Recently, it has been proposed that oxygen concentration may be one of those stimuli able to drive this transformation and thus for each oxygen level, there exists a dominant (fittest) tumor cell phenotype that corresponds to certain ratio proliferation/migration rates \citep{Giese}. Several studies have linked hypoxia to the invasive behavior of different types of tumors and their relationship with metastasis and negative prognosis \citep{Bristow,Kalliomaki,Elstner}. Mathematically, the observation that the hypoxic phenotype is more migratory than the normoxic one \citep{Berens,Giese,Bristow,Gorin} translates to choosing the hypoxic cell diffusion coefficient $D_h$ to be larger than the normoxic one $D_n$.
\par

The second terms in Eqs. (\ref{normox}) and (\ref{eqhipox}) employ a classical logistic growth for the tumor cell populations with proliferation times $\tau_n$ and $\tau_h$, respectively.  Thus, the net proliferation is lower in regions of high cell density where $C_n+C_h+C_d$ is close to the maximum density capacity $C^{\left(M\right)}$, than in regions of low cell density where $C_n+C_h+C_d$ is much smaller than $C^{\left(M\right)}$. As to the parameters $\tau_n$ and $\tau_h$, the migration-proliferation dichotomy suggests that $\tau_n < \tau_h$. Since growth is assumed to be space-limited we incorporate also the necrotic tissue density into the saturation terms.
\par

The third terms in Eqs. (\ref{normox}) and (\ref{eqhipox}) are essential ingredients of our model since they represent the phenotypic switch under the effect of changing microenvironmental conditions. The {\em switch} functions $S_{nh}$, $S_{hn}$ depend on the oxygen concentration (to be specified below) and account for the fact that under low oxygen conditions normoxic cells change their phenotype to the hypoxic one with a rate $1/\tau_{nh}$, whereas when exposed to sufficiently high oxygen concentrations hypoxic cells recover their oxic phenotype with a rate $1/\tau_{hn}$.
\par

Equations (\ref{themodel}) also include hypoxic cell loss by death due to anoxia [cf. last term in Eq. (\ref{eqhipox})] and how these lost cells feed the necrotic tissue [cf. Eq. (\ref{death})]. The last term in Eq. (\ref{eqhipox}) represents (anoxic) cells dying when the oxygenation conditions are adverse enough as measured by the switch function $S_{hd}$, with a rate $1/\tau_{hd}$. It should be emphasized  that cell loss in glioma occurs both by apoptosis (representing death by controlled suicide) and by necrosis (death by uncontrolled damage). Necrosis has historically been considered to be an inappropriate or accidental death that arises under conditions that are extremely unfavorable such as those incompatible with a critical normal physiological process. More recently, a number of studies have suggested that necrosis is also a regulated process that can be modulated. For example, induction of necrosis seems to be dependent on cellular energy stores, such as NAD and ATP. Furthermore, cell stress and cell signaling including oxidative stress, calcium levels and p53 activation have been shown to influence lysosomal membrane permeability and survival \citep{Boya}. When apoptosis sets in, the cell membrane collapses and the cell shrinks, while when necrosis occurs the cell keeps its shape and thus still occupies a physical space \citep{Hotchkiss}. This leads to the fact that cell loss due to apoptosis, not leading to the fill up of space, can be simply incorporated into effective proliferation cell rates in Eqs. (\ref{normox}) and (\ref{eqhipox}). However, cell death by necrosis happens at a different rate and, in addition, results in occupied space; this is the reason for incorporating the population $C_d$ into the proliferation limiting terms in Eqs. (\ref{normox}) and (\ref{eqhipox}). Overall, levels of apoptosis are generally low in malignant gliomas; apoptosis accounting for only a slim minority of cell death. Furthermore, apoptotic rates do not correlate with prognosis \citep{Migheli,Schiffer}. In contrast, coagulative necrosis represents the majority of cell death in GBMs and its degree is inversely related to patient survival, as several  shown by \citet{Nelson} and \citet{Lacroix}. 
\par

\subsection{Microenvironment oxygenation}
%\label{Oxygenconcentration}

Though the GBM microenvironment is highly heterogeneous at the cellular and molecular levels \citep{Bonavia}, two of the main chemical agents implicated in its growth and metabolism are oxygen and nutrients, mainly glucose and lactate \citep{Berta,Griguer05,Griguer08,Seyfried}. A more comprehensive description of the GBM microenvironment should, a priori, take into account the interplay not only of oxygen but of nutrients as well, at the expense of increasing the complexity of the mathematical model. Here we show that {\em it suffices} to take oxygen as the key chemical agent driving the collective cell migration dynamics to understand pseudopalisade formation in GBM. Moreover, glucose is typically less scarce than oxygen even at long distances from the blood vessels and can be replaced by other fuels such as lactate or, in a smaller fraction, by pyruvate, malate or glutamine \citep{DeBerardinis,Berta,Beckner,Grillon}. For instance, sequestration of lactate and pyruvate in the extracellular matrix has been observed in C6 gliomas in a rat brain by \citet{Grillon} and correlates with the persistent acidity of the GBM microenvironment.

\par
 
Oxygen tissue heterogeneities are recognized to be a very relevant factor in gliomas and many other tumor types \citep{Evans}. The spatio-temporal variation of oxygen distribution, driving the presence of various phenotypic tumor subpopulations, would help to explain the diversity of responses obtained from the same treatment employed in patients with apparently similar tumors but irrigated in a different way. Here, a Michaelis-Menten type kinetics is used for the uptake of oxygen \citep{Patel,Ferreira}, but now encompassing the feedback by the normoxic and hypoxic cells
\begin{equation}
	\frac{\partial O_2}{\partial t} =D_{O_2} \nabla^2 O_2 -  \frac{\alpha_n C_n + \alpha_h C_h}{O_2^{\left( T \right)} + O_2}O_2. \label{oxygen}
\end{equation}
The first term in Eq. (\ref{oxygen}) accounts for the oxygen diffusion in the brain tissue assuming a homogenous and isotropic diffusion coefficient $D_{O_2}$ for simplicity. Although oxygen passes successively through the intracellular fluid, cell membranes and cytoplasm, all having abrupt spatial variations, 
 employing an average diffusion coefficient for oxygen has been proven to be a good approximation in previous works \citep{Tannock,Pogue}. 
The second term models the oxygen consumption by normoxic  and hypoxic cells at rates $\alpha_n$ and  $\alpha_h$, respectively.  
The saturation Michaelis-Menten constant $O_2^{(T)}$ corresponds to the oxygen concentration level at which the reaction rate is halved. 
\par

\begin{figure}[t]
\begin{center}
%\vspace*{-2mm}
%\hspace*{1mm}
\includegraphics[scale=0.45]{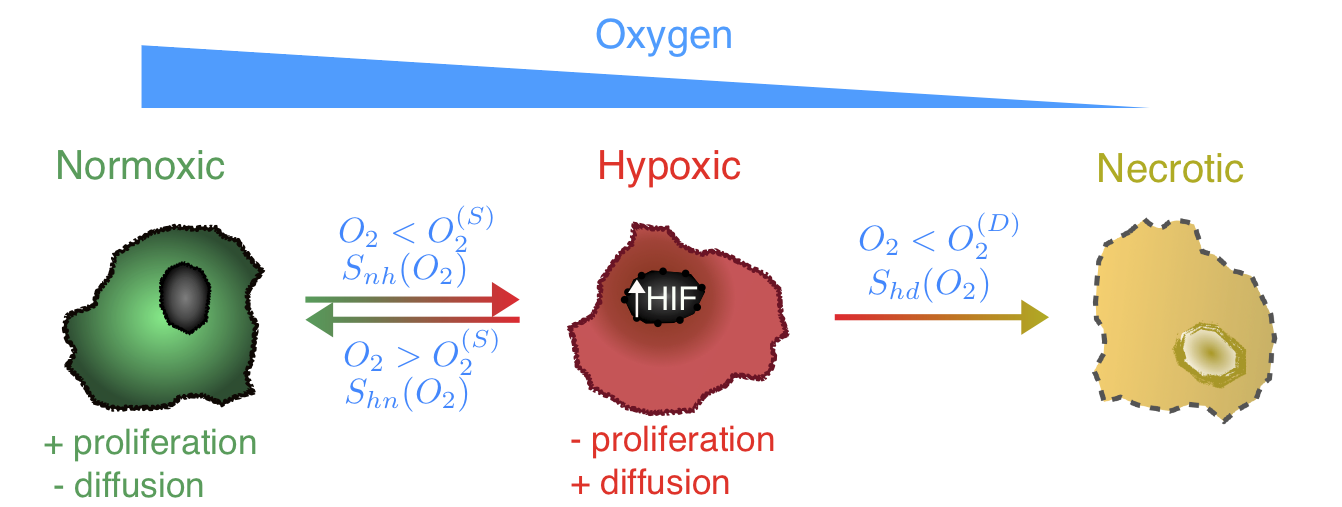} 
\end{center}
\caption{{\bf Oxygen concentration influences the phenotype of a tumor cell.} Depending on the oxygen concentration various switching mechanisms arise coupling the populations: Normoxic to hypoxic, $S_{nh}$, hypoxic to normoxic $S_{hn}$ and hypoxic to necrotic $S_{hd}$. High oxygen levels favor the existence of more proliferative phenotypes which are less mobile. On the contrary, cells respond to low oxygen concentrations by expressing
less proliferative phenotypes which are more motile, and finally, hypoxic cells experiencing persistent anoxia will eventually die \citep{pseudopalisading1}. \label{fig:presentation}}
\end{figure}

When modeling the oxygenation dynamics, we will assume that normoxic cells switch to the hypoxic phenotype with a characteristic time $\tau_{nh}$ when the local oxygen concentration is below a critical value $O_2^{(S)}$, such that hypoxia-inducible factor 1$\alpha$ (HIF-1$\alpha$) is not degraded and the glycolytic and angiogenesis mechanisms initiate \citep{Jewell}. The switching of cells from the hypoxic to the normoxic state $S_{hn}$, as described by Eqs.  (\ref{normox}) and (\ref{eqhipox}), occurs with a characteristic time $\tau_{hn}$ and is a reversible process $S_{nh}$; if the local oxygen sensed by a hypoxic cell is above $O_2^{(S)}$, the cell will undergo a phenotypic switch to the normoxic state (as it satisfies $O_2>O_2^{(S)}$) and viceversa. Finally, hypoxic cells experiencing persistent anoxia will eventually die when the oxygen concentration is below a threshold value $O_2^{(D)}$ with a characteristic time $\tau_{hd}$. The specific form of the three step-like switch functions are
\begin{subequations}
\label{switches}
\begin{eqnarray}
S _{hn} (O_2) & = & \frac{1}{2}\left[1+\tanh\left(\frac{O_2-O_2^{(S)}}{\Delta O_2}\right)\right], \label{sHN} \\
S _{nh} (O_2) & = & \frac{1}{2}\left[1-\tanh\left(\frac{O_2-O_2^{(S)}}{\Delta O_2}\right)\right], \label{sNH} \\
S _{hd} (O_2) & = & \frac{1}{2}\left[1-\tanh\left(\frac{O_2-O_2^{(D)}}{\Delta O_2}\right)\right]. \label{sHD}
\end{eqnarray}
\end{subequations}
The extra parameter $\Delta O_2$ accounts for the fact that transitions occur around the critical values $O_2^{(S)}$ and $O_2^{(D)}$ with a characteristic range (or width). In our calculations we have taken the same small width  $\Delta O_2$ for all of the transitions. The change from the hypoxic state to necrosis is irreversible and takes place in an interval centered around $O_2^{(D)}$, the size of this transition domain being proportional to $\Delta O_2$.

Although it is possible to incorporate the oxygen supply by resorting to extra source functions, here we will use a more direct approach and oxygen will be provided initially through the two capillaries located at the boundaries of the tumor domain (see Figure \ref{fig:scheme}). As soon as one of the two blood vessels becomes impaired by a thrombus, the remaining functional one will be the sole source of oxygen.
\par

\subsection{Parameter estimation}
 
We resort to available experimental values from human glioma models to obtain order-of-magnitude estimates of the intervening parameters in our equations. First, the maximum cell density $C^{(M)}$ has been estimated in previous works \citep[see e.g.][]{Swanson} to be of about $10^6$ cell/cm$^2$. The oxygen concentration at which cells switch to hypoxic metabolism depends on the specific cell line but experimental evidences support for glioma the choice of 7 mm Hg \citep{Vaupel2}. The Michaelis-Menten constant has to be smaller than this value yet larger than the anoxia threshold of about 0.7 mm Hg \citep{Brown}. We have chosen it to be about $O_2^{(T)} = 2.5$ mm Hg \citep{Das,David}. 
\par

For the oxygen consumption parameters, although there is a high variability between different cell lines, we will take the values of U251 glioma cells from \citet{Griguer08}. As to the hypoxic oxygen consumption, this is a parameter difficult to assess. We will use data for the same cell line U251 but genetically modified to deplete its mitochondrial DNA so that the cell's oxidative capacity is virtually impaired. Data from \citet{Griguer08} indicates that the normoxic cell uptake is five times higher than the hypoxic cell consumption. We have indeed assumed that the consumption rates were reported for normal conditions of pressure and temperature. Moreover, it is known that the mean oxygen pressure in arterial blood is around 95 mm Hg \citep{Kimura}, while venous values are around 30-40 mm Hg \citep{David} thus, we will take the oxygen pressure at our boundary capillaries to be $O_2^{v_i} =$ 60 mm Hg.
\par
The oxygen diffusion coefficient is classically known to be around $10^{-5}$ cm$^2$/s \citep{Mueller-Klieser1984} while the cell diffusion coefficients are not so readily accessible to obtain {\em in vivo}. We resort to the data from \citet{Wang}, were the authors have exploited a recently developed tool for quantifying glioma net invasion rates in individual patients using routinely available magnetic resonance images (MRI) from 32 patients to derive a net measure of the invasive capacity of the overall dominant phenotype. Since they obtain the parameters for the invasive phenotype, which is the most migratory one, and this is linked to the hypoxic one in our model, we will take their median migration rate 29 mm$^2$/year as our {\em fast} diffusion coefficient for hypoxic cells. The normoxic cell diffusion will be taken to be smaller than the hypoxic one, in our case $D_n= 0.1 D_h$.
\par
Finally, the proliferation parameters can be estimated from typical doubling times that range from 24 to 48 h for glioma cells in vitro for well oxygenated cells, while the mobile hypoxic cells doubling times are assumed to be longer (we have taken this parameter to range within two to four times the previous one). Table \ref{parameters} summarizes the typical parameter values employed in our calculations.

\begin{table}[h]
\caption{Values of the biological parameters used into our model equations.}
\label{parameters}
%\centering
\begin{tabular}{llll}
\hline\noalign{\smallskip}
Variable  & Description & Value & References \\
\noalign{\smallskip}\hline\noalign{\smallskip}
$C^{(M)}$& Maximum tumor  & $10^6$ cell/cm$^2$ & \citep{Swanson}\\ & cell density & & \\ \hline
$O_2^{(S)}$& Oxygen concentration & 7 mmHg & \citep{Vaupel2}\\ 
                 &  switch to hypoxia      &               &    \\ \hline
$O_2^{(T)}$&  Michaelis Menten  & 2.5 mmHg &{\citep{Das}}  \\% \citep{Tristan}\\ 
                 &  oxygen threshold   &   &  \\ \hline
 $O_2^{(D)}$& Oxygen level  & 0.7 mmHg & \citep{Brown}\\ & for anoxia & & \\ \hline
$O_2^{v_i}$$_{i=1,2}$& Vessel oxigen & 60 mmHg & \citep{David}\\
                                &   concentration                                         &                & \citep{Kimura}\\ \hline 
$\alpha_n$&Normoxic cell oxygen & $10^{-17}$  Mol/(c s) & \citep{Griguer08}\\  & consumption & & \\ \hline
 $\alpha_h$&Hypoxic cell oxygen & $\alpha_n/5$ Mol/(c s) & \citep{Griguer08}\\ & consumption & & \\ \hline
$D_n$& Diffusion coefficient for  & 6.6$\cdot$10$^{-12}$cm$^2$/s & \citep{Tjia}\\
          &      normoxic cells                                                   &                                                   &\citep{Wang} \\ \hline
$D_h$& Diffusion coefficient for  & 10 $D_n$  cm$^2$/s & \citep{Tjia}\\
          &         hypoxic cells                                                &                                                   &\citep{Wang} \\ \hline
$D_{O_2}$& Oxygen diffusion  &$10 ^{-5}$  cm$^2$/s & \citep{Mueller-Klieser1984} \\ & coefficient & & \\ \hline
$\tau_{nh}$&Phenotype switch time & 1 h & \citep{Jewell} \\
                &  (normoxic to hypoxic) & & \\ \hline
$\tau_{hn}$&  Phenotype switch time& 96 h & Estimated \\
& (hypoxic to normoxic) & & \\ \hline
$\tau_{hd}$ & Cell death mean & 48 h & Estimated \\ & time in anoxia & & \\ \hline
$\tau_n$ &Normoxic cell  doubling & 24 h&  \citep{Giese} \\
          &  time                                                         &                                                   &\citep{Berens} \\ \hline
$\tau_{h}$&Hypoxic cell doubling  & 48 h & \citep{Giese}\\
          &    time                                                     &                                                   &\citep{Berens} \\ \hline
$\Delta O_2$& Sensitivity around  & 0.1 Mol/mm$^3$  &Estimated \\ 
& transition threshold & & \\ \hline
$J$& Oxygen exchange   & 0.1 cm  & \citep{Mazumdar} \\ & coefficient & & \\
\noalign{\smallskip}\hline
\end{tabular}
\end{table}

\subsection{Computational details}
\label{computacion}

The model given by Eqs. (\ref{themodel}) and (\ref{oxygen}) describes the dynamics of cells in the bulk of the tumor under variable oxygenation conditions. As such, it can be studied on any particular geometry either regular or irregular with suitable boundary conditions and/or oxygen sources. It can also be coupled to detailed models of tumor vasculature such as those based on phase-field models \citep{Travasso} and/or combined with therapies having an oxygen-dependent effectiveness such as radiotherapy \citep{Jensen,Vaupel,Bristow,Wilson}.

\par

Here, we wish to analyze our model to get insight on the formation of necrotic areas and hypercellular regions such as those observed in gliomas. To do so, we will stick to the simplest possible scenarios: either two dimensional rectangular tumor geometries with vessels located on two opposite boundaries and periodic boundary conditions on the other (opposed) two sides, or just one dimensional sections of the previous system with oxygen flow coming from the boundaries (end points) of a one-dimensional domain. 
\par

The sources of nutrients and oxygen in the brain are the blood vessels which, in our 2D model, correspond to two of the four boundaries of the tumor tissue analyzed and, in the 1D case, to two sides of the interval under study. Here, the blood vessels are assumed to be capillaries.
% (microscopic single-thickness vessels that connect the arterial and venous segments of the circulatory system). 
Exchange of gases, nutrients, and waste materials occurs through the thin permeable walls of these capillaries. We will denote the oxygen concentration on the two lateral blood vessels at a given time $t$ by $O_2^{v_1}(t)$ and $O_2^{v_2}(t)$. Oxygen flows from vessels to the tissue to balance the different oxygen pressures, thus the boundary conditions for the oxygen at the blood vessels are 
\begin{equation}
	\frac{\partial O_2}{\partial x} =\frac{1}{J_i} (O_2^{v_i}(t) - O_2), \quad  i=1,2. \label{bc2}
\end{equation}
The capillary length in our model simulations is taken to be $\leq 1$ mm to match the real capillary size  \citep{Mazumdar}. In addition, oxygen concentration in the capillary network does not follow the variations induced by the circulation in the major blood vessels \citep{Das}.  Since our computational domain extends over a small spatial region ($\simeq$700 $\mu$m in length parallel to the blood vessels), we will assume the oxygenation to be the same independently on the position along the capillaries as our computational domain extends over a small spatial region (a strip of about 700 $\mu$m in length parallel to the blood vessels).
\par

Likewise, the oxygen exchange coefficient $J_i$ is considered to be constant and equal for both vessels, therefore $\vert J_1\vert=\vert J_2\vert=J$. Bearing in mind that the average blood velocity for capillaries is around $0.1$ cm/s  \citep{Mazumdar} we can estimate its value to be $J \sim$ $0.1$ cm. Notice that our boundary conditions (\ref{bc2}) imply that there exists the possibility that oxygen spreads either away from or into the vessel should the tissue oxygen concentration be lower that or exceed the oxygen pressure at the vessel.
\par
Once the computational domain has been fixed, we impose the boundary conditions for the cell densities to supplement model equations (\ref{themodel}) and (\ref{oxygen}). The following three equations represent the flux conditions describing cell intravasation into the blood vessels
\begin{subequations}
\label{bc}
\begin{eqnarray}
\label{borderN}
	\frac{\partial C_n}{\partial x}  & =  & -\frac{1}{J_n} C_n, \\
\label{borderH}
	\frac{\partial C_h}{\partial x} & = & -\frac{1}{J_h} C_h, \\
	\frac{\partial C_d}{\partial x} & = & 0,
\end{eqnarray}
\end{subequations}
Note that it is possible to calculate the number of cells that intravasate into the blood flow for $t\in [0,T]$, $T$ being the studied temporal window, by means of the formula
\begin{equation}
	M(t)=\int_0^T\left(J_n C_n + J_h C_h\right) \,dt ,
\end{equation}
where, in general, it is to be expected that the hypoxic intravasation coefficient $J_h$ would be substantially larger than the normoxic one $J_n$, due to the higher motility of the former cells. In metastatic processes, solid tumor cells enter into the lymphatic and blood vessels eventually leading to the formation of colonies in distant organs, thus Eqs.  \eqref{borderN} and \eqref{borderH} represent the first stage of metastasis. However, GBM only rarely metastatize out of the brain. Although GBM cells may spread along blood vessel walls, primary brain tumors do neither intravasate to blood vessel walls \citep{Bellail} nor invade into the bony calvarium which encloses the brain.
Indeed, the capacity for vascular invasion is a primary requisite for wide hematogenous metastasis of malignant neoplasms. Thus, the absence of vascular invasion
by primary brain tumors, taken together with the absence of lymphatics within the brain, may explain the observation that primary brain tumors only very rarely metastasize outside of the brain to ectopic organs. The unique patterns of invasion and metastasis shown by primary brain tumors suggest that these neoplasms are adapted to the microenvironment specifically present within the central nervous system.
\par
Therefore, we will assume henceforth that for gliomas $J_n = J_h = 0$. However, our model equations together with Eqs. (\ref{bc}), can be used in more general contexts, e.g. to elucidate the relation between acute hypoxia or cyclic hypoxia and metastatic processes in other types of tumors. 
\par
To solve our model set of equations with the boundary conditions on the capillaries, plus periodic boundary conditions on the free sides, we have used standard finite differences (second order in space) either with an explicit fourth order Runge-Kutta method in time or an implicit second order in time Crank-Nicholson-type scheme. Typical parameter values for the spatial step were $\Delta x= \Delta y$= 5  $\mu$m, and  $\Delta t$ = 0.01 s for the explicit scheme. 

In all of the simulations to be presented later we thus assume an initial concentration of normoxic tumor cells in the region limited between two functional vessels. Thus, initially the 
hypoxic and necrotic cell densities are zero for all $x \in (0 , L_{x})$. The initial oxygen distribution is randomized with 24 mm Hg as the mean oxygen partial pressure as described by \citep{MeanpO2}.
Except when otherwise stated, we will assume that the vessel $v_1$ is disrupted at $t=0$ so that $v_2$ remains as the only functional vessel through the simulation. \par

\section{Results}
\label{results}

\subsection{Vaso-occlusive events trigger necrosis, the formation of pseudopalisades and vessel-cooption}

We have solved our model equations (\ref{themodel}) and (\ref{oxygen}) by including only the essential biological facts at the cellular level allowing us to understand the formation of palisading necrosis, and to get an insight on the sequence of events leading to their formation as well as estimates of their lifetimes and sizes. Once the parameters are chosen in the appropriate realistic ranges (see Table \ref{parameters}), the formation of pseudopalisades does not require a detailed fine-tuning of the parameters but it is a characteristic phenomenon arising in many of our simulations. It is interesting to highlight that pseudopalisades are transient states, being part of a sequence of events to be summarized in what follows.

\par

\begin{figure}%[h]
\begin{center}
\includegraphics[scale=0.58]{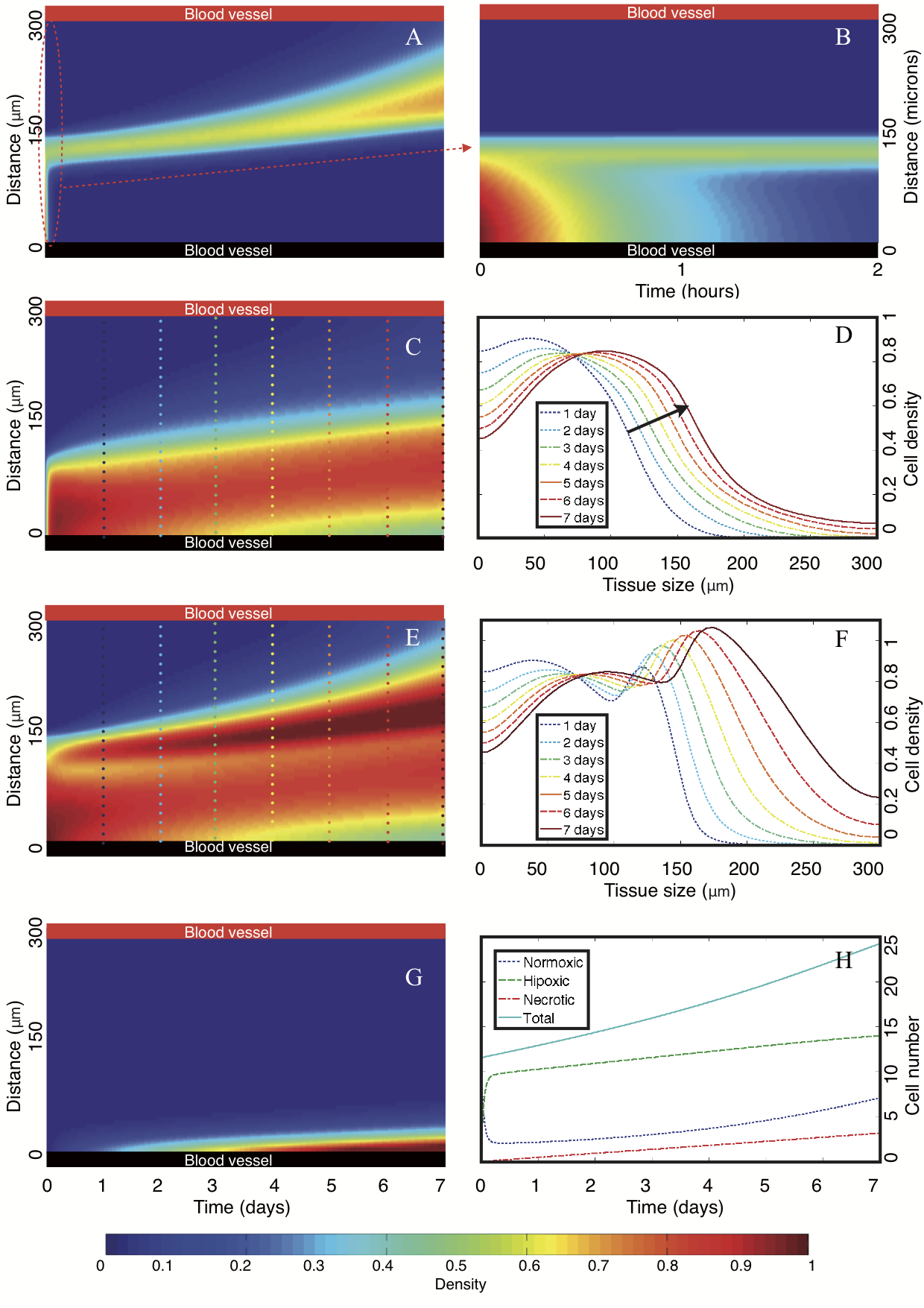} 
\caption{\textbf{Spatio-temporal simulations showing the formation of necrotic areas surrounded by hypercellular regions (pseudopalisades).} Initially, normoxic GBM cells are seeded close to the lower vessel. Pseudo color plots in A, B, C, E and G represent normoxic, normoxic (notice the scale), hypoxic, total (hypoxic +normoxic) populations and necrotic tissue densities, respectively. The horizontal and vertical axes correspond to time in days (except for B in hours) and space in $\mu$m, respectively. D and F depict the normoxic and total (normoxic + hypoxic) cell densities as a function of space for specific times indicated in each curve and also with vertical color lines in C and E. H shows the GBM cell number evolution with time for the different populations. At $t=0$, the lower vessel suffers a thrombus, while the upper vessel remains fully functional during the simulation. The parameters used in the simulations are listed in Table \ref{parameters}.}
\label{Densityplots1}
\end{center}
\end{figure}

In Fig. \ref{Densityplots1} we present a first scenario of the results obtained when normoxic GBM cells are seeded close to one of two vessels located at the boundaries of a brain tissue slice of 300 $\mu$m. Initial normoxic tumor cell density is maximum around one of the blood vessels (colored black in the lower side of Fig. \ref{Densityplots1}). The initial conditions for this simulations are taken to be of the form $C_{n}(x,0) = \sqrt{1- x^2/(L_{x}/2)^2}$ , $x \in (0 , L_{x}/2]$  and $C_{n}(x,0)=0$  for $x \in (L_{x}/2 , L_{x}]$  with $L_{x}=$300 $\mu$m.
 We assume that the brain tissue is well oxygenated until the lower vessel is disrupted at $t=0$  so that, at the outset, the hypoxic and necrotic cell densities are zero in the whole computational domain. The results for the evolution of the different cell densities are summarized in Fig. \ref{Densityplots1}. 
 
\par

After the lower blood vessel undergoes a thrombus induced by the adjacent high density of proliferating  normoxic tumor cells, these cells experience an oxygen shortage and  change their phenotype to hypoxic. This leads to an enhanced mobility and reduced proliferation in the perivascular region in an attempt to find a more nourishing environment to survive (Fig. \ref{Densityplots1}A). A complete phenotypic switch occurs in about 1 hour (see Fig. \ref{Densityplots1}B) so that the new hypoxic cells (Fig. \ref{Densityplots1}C) increase their flux to the upper vessel (in red) that is still functional, generating in the process a transient traveling wave of hypoxic cells (see Fig. \ref{Densityplots1}D where the arrow shows the direction of the movement). 
\par

\begin{figure} [h]
\begin{center}
\includegraphics[width=0.8\textwidth]{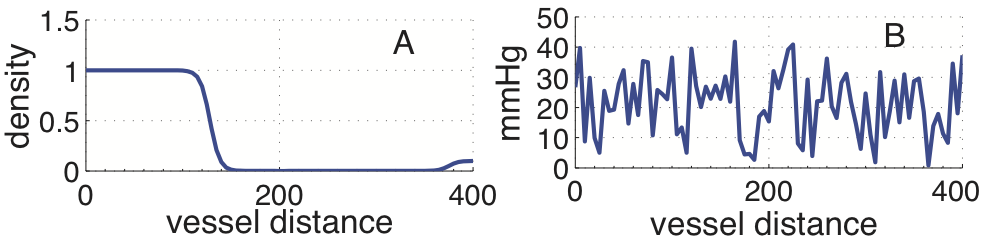} 
\caption{\textbf{Initial conditions for the simulation displayed in Fig. \ref{Densityplots2}.} (A) Normoxic cell density.  (B) Initial oxygen distribution with 24 mmHg as the average oxygen pressure for a healthy brain tissue \citep{MeanpO2}.} 
\label{initial_conditions}
\end{center}
\end{figure}
Figs. \ref{Densityplots1}(E,F) depict the formation of a pesudopalisade in the scale of 1 day that develops during the subsequent days. A morphologic analysis of Figs. \ref{Densityplots1}E and F gives about 130 $\mu$m as the characteristic width value. According to  \citet{pseudopalisading1}, pseudopalisades with small widths have smaller necrotic cores. In our case, the necrotic core is about 30 $\mu$m (Fig. \ref{Densityplots1}G) and it is also formed in the same time scale of days.

\begin{figure}
\begin{center}
%\vspace*{-2mm}
%\hspace*{1mm}
\includegraphics[scale=0.5]{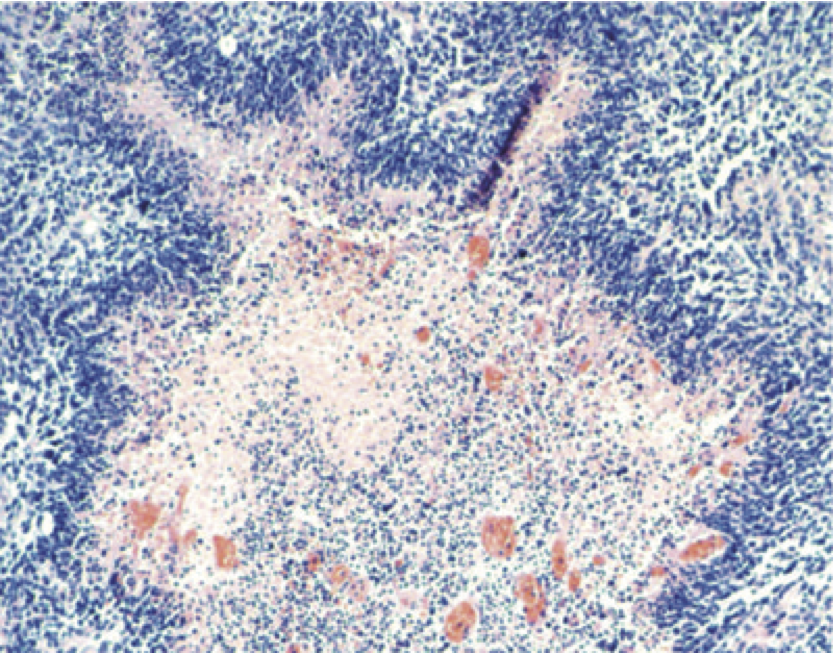} 
\end{center}
\caption{{\bf Vascular damage as a driving force for migration in Glioblastoma Multiforme.}
Formation of necrosis and the emergence of pseudopalisades is part of a series of related events that are crucial for the accelerated progression of high grade astrocytic gliomas. The development of an envelope of small scale pseudopalisades may play a relevant role in the global migratory dynamics and invasion of these malignant gliomas driven by vaso-occlusive events occurring at the small cellular scale, leading to the distinctive feature in GBM of showing significant necrotic areas where frequently central vessels are present \citep{pseudopalisading1}. \label{fig:largepalisade_Brat}}
\end{figure}

Under persistent anoxic conditions, hypoxic cells located around the impaired vessel start to die (Fig. \ref{Densityplots1}G). Nevertheless, there is a significant density of migrating hypoxic cells that reach the functional vessel.  Finally, Fig. \ref{Densityplots1}H illustrates the time evolution of the total GBM cell number for each population in the full simulated tissue. It is observed that hypoxic cell number increases significantly fast during the first hours after the thrombosis event but then moderates its growth rate. Normoxic cell numbers have a different growth pattern with an initial decay and a final relapse, once the functional vessel has been co-opted. So, in less than a week, tumor cells have invaded the 300 $\mu$m of tissue. 

\par

\begin{figure}%[h]
\begin{center}
\includegraphics[scale=0.58]{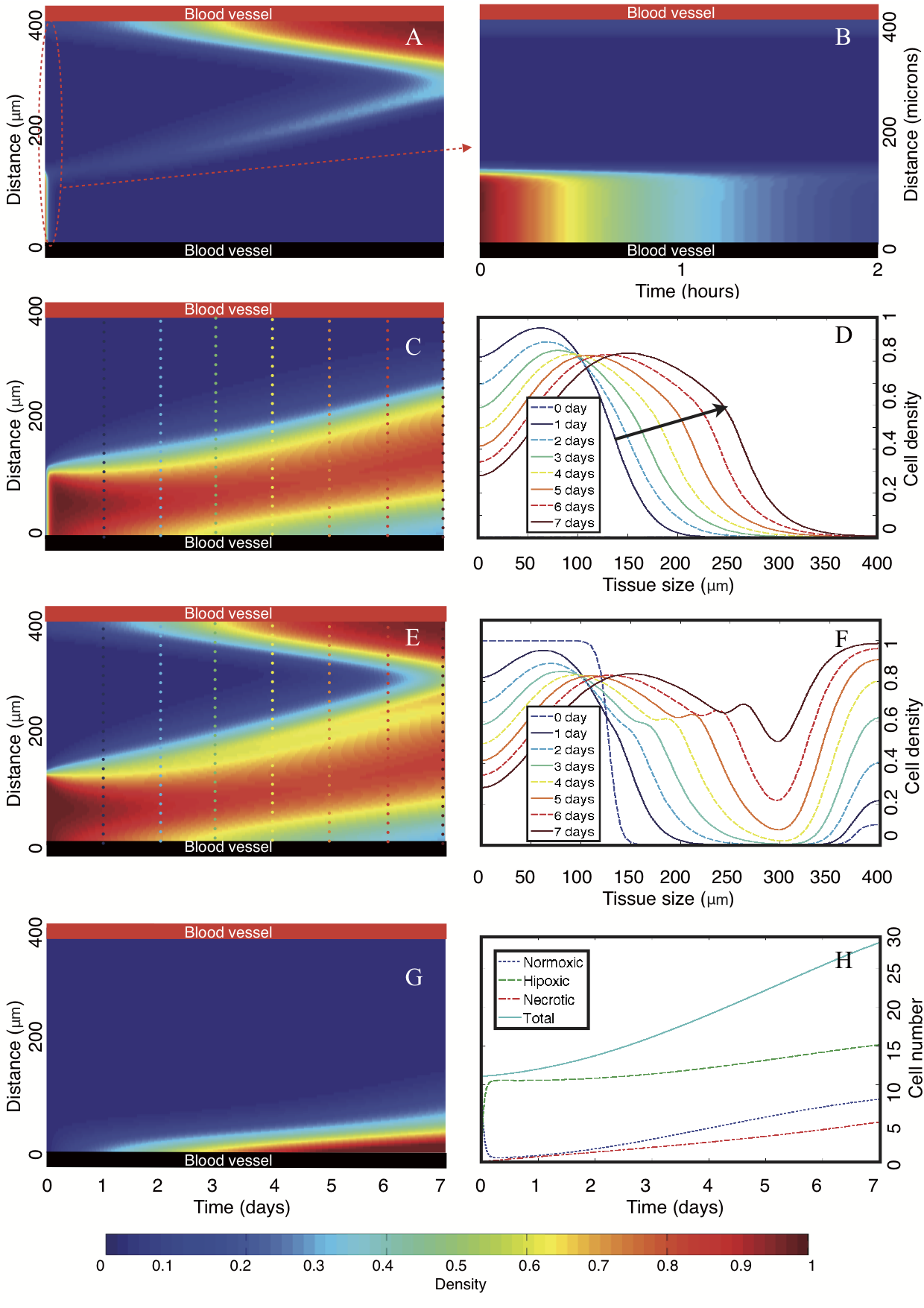} 
\caption{\textbf{Spatio-temporal simulations showing the formation of perinecrotic pseudopalisades after a vaso occlusive event.} In contrast with Figure \ref{Densityplots1}, here both vessels are initially seeded with normoxic GBM cells. After $7$ days, the upper vessel is completely invaded by a high GBM cell density and again a vaso occlusion event may occur at any moment. All of the subplots and the rest of parameters are as in Figure \ref{Densityplots1} and the initial data for oxygen and tumor cells are shown in Fig. \ref{initial_conditions}.}
 \label{Densityplots2}
\end{center}
\end{figure}

In a second set of simulations, displayed in Figure \ref{Densityplots2}, we have studied the invasion of a tumor vessel by a pseudopalisading wave of hypoxic cells coming from another impaired vessel. As in the previous case, a  high normoxic GBM cell density is seeded around the lower blood vessel (see  Fig. \ref{Densityplots2}), but now we assume that a small number of tumor (normoxic) cells have already migrated to the upper blood vessel starting to colonize the second vessel (400 $\mu$m distant from the lower vessel). The initial conditions for the tumor (normoxic) cell density and oxygen concentrations are shown in Fig.  \ref{initial_conditions} and for the normoxic tumor cell density are given by $C_{n}(x,0)=\frac{1}{2} (1-\tanh(\frac{x-L_{x}/3}{2}))$, $x \in (0 , L_{x}/2]$ and $C_{n}(x,0)=0.05 (1+\tanh(\frac{x-L_{x}-20}{2}))$  for $x \in (L_{x}/2 , L_{x}]$  with $L_{x}=$400 $\mu$m.

The simulation starts when the lower vessel suffers a thrombus induced by the adjacent high density of proliferating  normoxic tumor cells. The upper vessel is fully functional during the simulation. This vessel is completely invaded, after 7 days, by a high tumor cell density that might induce new vaso-occlusive events (Fig. \ref{Densityplots2}A). The new hypoxic cells (Fig. \ref{Densityplots2}C) generate a transient traveling wave (Fig. \ref{Densityplots2}D) where the arrow shows the direction of the movement. Figures \ref{Densityplots2}E and F show the formation of a pseudopalisade in the time scale of one day that develops during the ensuing 7 days.

A morphologic analysis of Figs. \ref{Densityplots2}E and F gives about 150 $\mu$m as the characteristic width value.  The necrotic core is about 60 $\mu$m (Fig. \ref{Densityplots2}G) and it is also formed in the same time scale of days. Finally, it is observed in Fig. \ref{Densityplots2}H that the hypoxic cell number increases significantly fast during the first hours after the thrombosis event but then moderates its growth rate. Normoxic cell numbers have a different growth pattern with an initial decay and a final relapse, once the functional vessel has been co-opted.
\par   

Qualitatively, it is important to underscore that this minimal model not only reproduces the clinical observations seen in immunohistochemical analysis of histologic sections of GBM samples \citep{pseudopalisading1}  but may also contribute to a better understanding of the origin and evolution of those structures and to find unknown quantities such as the typical palisade lifetime or the relationship between palisade lifetime/size and vessel distance. Moreover, the macroscopic progression of GBM can be partially conceived, from a microscopic point of view, as consisting of a large number of intravascular thrombotic events (caused by unregulated tumor cell proliferation in the vicinity of the vessels) giving rise to the formation and coalescence of small necrotic foci coupled with subsequent episodes of hypoxic cell migrations in search of nearby functional blood vessels which will eventually suffer new intravascular thrombotic events. The envelope of many of these small-scale pseudopalisades contributes to the high density regions on larger spatial scales as those shown in Fig. \ref{fig:largepalisade_Brat}.

\subsection{Palisading waves leads to invasion faster than that of pure random motion invasion}

In principle, it is not clear a priori how the vaso-occlusive events relate to the speed of the tumor invasion. Phenotypic characteristics of GBM such as hypercoagulation, hypoxia, and abnormal angiogenesis, may be linked at the molecular level, and hypoxia may coordinate the hypercoagulative activity of GBM cells and (protease-activated receptor) PAR-mediated angiogenic signaling \citep{Svensson}.

\par

On the one hand, one might naively argue that the vessel maintenance would lead to an increased number of cells that by random motion (diffusion) would give rise to a faster 
invasion due to a higher tumor cell number. On the other hand, one might also argue that vaso-occlusive events, despite resulting in hypoxia, cell death and slower proliferation, could induce more mobile cells and might instead accelerate the overall GBM progression. To gain further quantitative insight of what is the real situation, i.e. what is the role of vaso-occlusive events in relation with the tumor progression speeds, we have performed a series of numerical simulations.

\begin{figure}%[h]
\begin{center}
\includegraphics[scale=0.55]{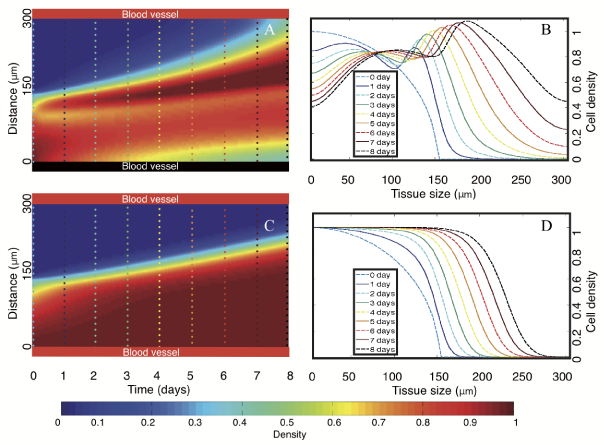} 
\caption{\textbf{Palisading waves lead to invasion faster than that by pure random motion invasion. } (A-B) Evolution of the tumor cell densities as a function of time and space when only the upper vessel is functional while the lower vessel becomes impaired at $t=0$. Parameters are as in Fig. \ref{Densityplots1}.  Pseudocolor plot A depicts the total GBM cell density as a function of time, while plot B presents density curves at various times showing the advance of the tumor front. 
(C-D)  Same as in plots (A-B) but when both vessels remain functional. All parameters and initial conditions are taken as in Figure \ref{Densityplots1}.}
\label{Heparina}
\end{center}
\end{figure}

Fig. \ref{Heparina} is a representative example of our results. Subplots A,B in Fig. \ref{Heparina} correspond to the case discussed previously where the upper vessel works normally but the lower vessel becomes impaired due to the formation of a thrombus. On the contrary, Fig. \ref{Heparina}C,D shows the tumor progression when both vessels remain functional.  In the first case, the functional vessel is invaded by the tumor after $8$ days by a high tumor cell density (more than 45$\%$) however when no thrombus is formed and both vessels stay functional (subplots C and D)  invasion becomes significantly slower; requiring $13.5$ days to attain the same tumor cell density around the upper vessel. Fig. \ref{Heparina} displays the coexistence of two mechanisms driving invasion: the first one being purely diffusive and similar to that governed by a free growth (when resources are not limited) obeying the Fisher-Kolmogorov equation, as described by \citep{V2} and related works. The second one shows an accelerated invasion caused by micro thrombi formations in the tumor vessels. Thus, on the basis of our model, it appears that vaso-occlusive events in GBM may play a key role in accelerating the tumor invasion process through pseudopalisade formation. 

Thus our model simulations suggest that the use of chemical agents slowing-down the vessel impair might delay tumor progression in GBM patients. This fact will be discussed throughly in Sec. \ref{sec_discussions}.

\subsection{There is a relation between the size of broken vessels, the palisade lifetime and the width of the necrotic region}

\begin{figure}%[h]
\begin{center}
\includegraphics[scale=0.30]{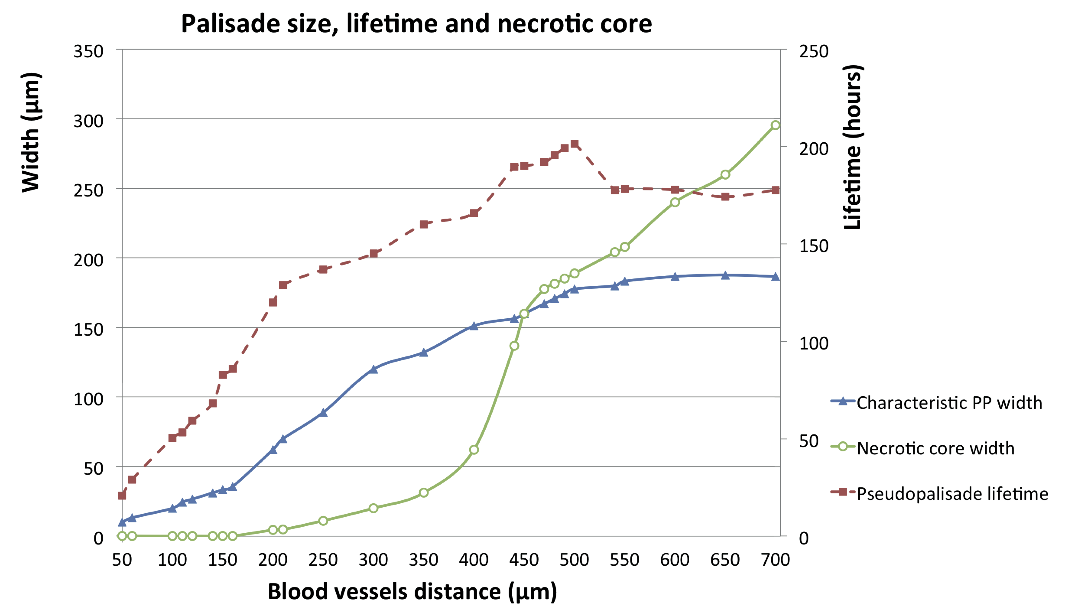} 
\caption{\textbf{Dependences between thrombotic-functional vessel separation, palisade width and its lifetime.} In all of our simulations GBM normoxic cells are seeded close to the impaired vessel with a small fraction of them already invading the functional vessel. Initial conditions are chosen similar to those of Fig. \ref{Densityplots2} and proportional to the tissue size. At $t=0$ one of the vessels becomes thrombotic while the second vessel remains fully functional during the simulated time window.  Characteristic pseudopalisade width and lifetime data are labelled by blue triangles, and red squares, respectively.  Also shown is the dependence of the necrotic core width with the tissue size (green circles). The horizontal axis represents the distance between blood vessels (tissue size) in $\mu$m. The vertical axes label (left one) the palisade and the necrotic foci widths in $\mu$m; (right one) the pseudopalisade lifetime in hours. The parameters used in the simulations are listed in Table \ref{parameters}.}
 \label{Lifetime}
\end{center}
\end{figure}

To further substantiate our results, we have carried out a numerical study of the relation between the separation between a thrombotic vessel and another functional vessel, the palisade width and its lifetime, as depicted in Fig. \ref{Lifetime}. The characteristic width of the pseudopalisade corresponds to the dimensions of the hypercellular region and its lifetime is the time elapsed since the palisade is formed until it disappears. Although the standard distances between blood vessels are in the range of 100 to 400 $\mu$m, due to the irregular and aberrant structure of the tumor vasculature, we have considered a wider range (from 50 to 700 $\mu$m) assuming that those histologic samples showing distances larger than 700 $\mu$m will be formed by cells for which their main oxygen source would probably come from a vessel close to them but not visible in pathological analyses due to its spatial arrangement. 

Although our analysis correspond to one-dimensional scenarios we still expect to get a hint on the main tendencies, since more complex higher-dimensional scenarios with a realistic three-dimensional vessel network are too complicated to be sistematized.

\par 

In our computational window, the distance between blood vessels equals the tissue size and so when this is smaller than 100 $\mu$m the formation of pseudopalisades is practically non-existent.  In addition, Fig.  \ref{Lifetime} shows how the pseudopalisade size depends on the distance between blood vessels (that may be also expected to be roughly proportional to the vessel size), growing in a linear fashion as the distance increases until it reaches a plateau value close to 200 $\mu$m. Notice that vessel separations larger than 500 $\mu$m give rise to pseudopalisades that do not experience a significant width change despite the fact that a smaller difference between the pseudopalisade cell density and the adjacent tissue cell density is observed in our calculations. Also, as it might be expected, there is no appreciable necrosis until the distance between vessels is larger than 150 $\mu$m. Then, in the range from 150 to 450 $\mu$m the necrotic core grows exponentially and from here its growth becomes linear (Fig.  \ref{Lifetime}). In other words, if the distance from a thrombotic vessel to the nearest functional vessel is short enough, the cells arrive to better oxygenated areas without experiencing severe hypoxic stresses (thus diminishing their death) and reducing the palisade lifetime. However, if this distance is larger than 150-200 $\mu$m (or if the broken vessels are larger), there will be a considerable number of hypoxic cells unable to scape from severe hypoxia  giving rise to larger necrotic areas and longer palisade lifetimes. Here again, the model although biologically simple and solved on a simple domain, allows to explain the observed feature that large pseudopalisades, i.e. pseudopalisades surrounding larger broken vessels are more persistent and thus easier to observe, than small ones (cf. Fig. \ref{Lifetime}). 

\par

\section{Perspectives and Therapeutic implications}
\label{sec_discussions}

\subsection{Perspectives}

The model presented in this work can be extended in many ways. However the addition of vasculature seems the more natural way providing the basis to construct a more complex {\em in silico} framework to apply the different therapeutic modalities in a more realistic manner once the underlying mechanisms involved in palisade formation have been laid down. One possibility is to maintain equations \eqref{normox}, \eqref{eqhipox} and \eqref{death} respectively for the normoxic, hypoxic and necrotic populations substituting the microscopic oxygen densities by macroscopic variables such as blood vessel densities or vessel functionality $v(x,t)$ that are to be related to the oxygenation conditions. This vessel density would evolve according to an equation of the form
\begin{equation} \label{vasculature}
\frac{\partial v}{\partial t}  =  \frac{1}{\tau_v}\text{VEGF}(C_h,C_d) - \frac{1}{\tau_t}\text{THROMBOSIS}(C_n).
\end{equation}
accounting for the fact that, given that space is available, hypoxic-migratory cells $C_h$ contribute to the generation of the new vasculature $v$ through the secretion of  vascular endothelial grow factors (VEGF) (first term in Eq. \ref{vasculature}) and that normoxic cells contribute to the vaso-occlusive events (second term in  Eq. \ref{vasculature}).
\par

These ideas might provide a simple way to incorporate palisade formation due to micro-vascular damage to  
 full-brain scale mathematical models of glioma progression \citep{V1,V2,V3,V4,V5,Swanson,V6}  may allow to construct more realistic ``macroscopic" models. These models might be better suited to describe more realistically glioma progression and incorporate the effect of therapies and as such to provide clinicians with valuable tools for treatment optimization. 
\par

Other possible extensions of our model regard the presence of a small number of treatment-resistance (cancer-stem) cells in glioblastoma multiforme (GBM) and whether their metabolism can be effectively targeted. Indeed, determining whether this small fraction mainly depends on aerobic glycolysis or on oxidative phosphorylation has recently been the subject of considerable interest \citep{Pistollato}. It is known that in glycolytic tumors, which is the case of high grade gliomas, oxidative phosphorylation is not completely shut down since it provides a significant amount of ATP even under low glucose requirements \citep{Vander,Mathupala}. In this work we have focused on oxygen as one key factor known to strongly affect GBM progression and response to treatment (e.g. radiotherapy). While our distinction of the tumor cell phenotypes has relied on their oxic state, we anticipate that, among the migratory hypoxic cells seen in the pseudopalisades, a small fraction of them is formed by treatment-resistance cells. Actually, our simulations suggest that during the time window where hypoxic cells migrate from an impaired blood vessel to a functional one any treatment critically depending on the presence of oxygen will exhibit a poor clinical response. The importance of oxygen in these subpopulations is further supported by recent experiments by \citet{Vlashi} in which it was observed that in various GBM cell lines both progenitor cells as well as those exhibiting a stem-cell-like behavior consume less glucose and produce less lactate while maintaining higher ATP levels than their differentiated progeny. Their conclusions were that GBM stem cells rely mostly on oxidative phosphorylation. Even from an evolutionary point of view, the collective migration of hypoxic GBM cells when forming pseudopalisades appears to be a selection mechanism of the fittest cell phenotypes. 
\par
It is necessary to underscore that perinecrotic pseudopalisades formation in GBM is not a minor detail or yet another layer to add to the complexity of high grade glioma growth. There is ample evidence that the formation of necrosis and the appearance of pseudopalisades is part of a series of related events that are crucial for the accelerated progression of high grade astrocytic gliomas. Thus, the development of an envelope of small scale pseudopalisades may play a relevant role in the global migratory dynamics and invasion of these malignant gliomas driven by vaso-occlusive events occurring at the small cellular scale, leading to the distinctive GBM feature of showing significant necrotic areas, as observed in clinical imaging as it is exhibited in Figure \ref{fig:largepalisade_Brat}. 
\par
On the one hand, this hystopatologic feature is a distinctive feature of high-grade gliomas and does not arise in other types of tumors, probably due to the very high mobility of tumor cells in astrocytoma and the low resistance to invasion by the brain parenchyma. On the other hand, it is well known that gliomas are very complex and heterogeneous and yet, they can be understood in terms of two cell subpopulations corresponding to two dominant phenotypes \citep{DeBerardinis,Giese,Keunen,Onishi}: a population of proliferative, less mobile cells located in perivascular areas and a second population of less proliferative and more mobile cells typically infiltrating the brain parenchyma.  This means that minimal mathematical models such as those based on Fisher-Kolmogorov type equations accounting for GBM progression \citep{V1,V2,V3,V4,V5,Swanson,V6} may benefit from incorporating the two different phenotypes and necrosis as it is done in this paper. 
\par
However, in order to incorporate necrosis not only phenomenologically, as in \citet{V6}, but including at least some underlying mechanisms, it is necessary to take into account the basic intervening steps involved in palisade formation. In fact, to allow for the vasculature growth in response to the pro-angiogenic factors and breakup due to vaso-occlusive events, blood vessel dynamics must be included in the model. Moreover, oxygen variations occur in very small spatial scales and at very fast time scales that cannot be accounted for in detail when solving the problem at larger full-brain scales and for times of the order of months. This means that more global models must rely on the higher scale effective vascular density or vascular capacity $m(x,t)$ rather than on oxygen. This is essential since one of the conclusions of 
 \citet{pseudopalisading1} and \citet{vaso-occlusive}, also supported by our simulations, is that vascular damage is not only a key event in the formation of necrotic areas but also a driving force for migration, and this dynamics is linked to the evolution of the cell densities that are responsible for the vessel destruction (normoxic cell population in the perivascular areas) and creation (hypoxic cell subpopulations secreting VEGF and other pro-angiogenic factors).
 \par

  \subsection{Therapeutic implications}
  
One of the main results of our work is the prediction that \emph{targetting vaso-occlusive events in GBM might delay the tumor progression}, a result with relevant potential implications since GBM is a type of tumor for which even modest survival increases are considered to be a big success because of its poor prognosis. 

\par

Thromboembolism, i.e. vaso-occlusive events of large vessels, is well recognized as a major complication
of cancer and known to be a common cause of death
in cancer patients. There is strong evidence linking venous
 thromboembolic events and malignancy. Laboratory markers 
 of coagulation activation such as
thrombinÐantithrombin complex or prothrombin fragments 
 1 + 2 support the premise that malignancy is a hypercoagulable state. 
 Inflammatory cytokines (e.g. tumor necrosis factor and interferon), coagulation proteins 
 (e.g. tissue factor and factor VIII), and procoagulant microparticles
may be elevated in patients with malignancy. However, the relative
contribution of chemotherapeutics, tumor cells, endothelium, and circulating 
 procoagulants in promoting thrombus formation continues to be
investigated \citep{Khorana,Furie,Zwicker,Green}.

\par

Thromboembolic complications include a
broad spectrum of clinical problems, a fact that has
lead to the use of
thromboprophylaxis in a variety of forms \citep{Dolovich,Khorana,Green} for cancer patients. Specifically, gliomas have a high incidence of venous
thromboembolism (VTE); with several studies suggesting that 25\%-30\% of these
patients sustain thromboembolic events \citep{Streiff,Simanek}. 
The fact that glioma cell lines secrete pro-thrombotic factors has been known for a long time \citep{Bastida} and it is also known that the more tumoral tissue is removed during surgery of high-grade gliomas the less-likely are the patients to die from VTE \citep{Brose,Simanek}.
This fact has led to the consideration of thromboprophylaxis for glioma patients, specially in subjects with higher potential risks of developing venous  thromboses \citep{Hamilton,Batchelor,Jenkins,Khorana}.

\par

In addition to the prevention of VTE there are several direct mechanisms of action of some anticoagulants, such as LMWH, on tumor cells: direct cell killing \citep{Santos}, anti-angiogenic effects \citep{Svensson} and many other \citep[see e.g.][Chap. 15]{Green}. The results from our model seem to imply that an additional indirect antitumoral effect from thromboprophyylaxis 
might be expected related to the delay of tumor invasion, different from the direct antitumoral effect or the increase of life expectancy obtained from the prevention of the formation of big coagulates. 

\par 

The main limitation in using LMWH in post-operative patients has been related to the possible increase in bleeding events, that in the case of brain surgery are potentially very harmful. However, a limited phase II clinical study of LMWH therapies for high grade glioma patients by \citet{Robins} has shown that the therapy is safe. A recent phase III trial with a small number of patients seems to confirm this trend \cite{Perry}.

\par

If, as predicted, preventing capillar thrombosis induced by the tumor, results in a growth delay for the tumor, a synergistic positive effect with radiation therapy is to be expected. Having a better oxygenation and reducing hypoxia would lead to an enhanced tumor radio-sensitivity, that might provide an extra benefit in survival.

\par

Whether these predictions are clinically relevant remains to be studied, first in animal models testing if the use of LMWH induces a significant reduction of the formation time for palisades and delays tumor progression and, in case of success, in more complete clinical trials. The fact that mathematical modelling might contribute to designing therapies against such an aggressive type of brain tumor is encouraging and we expect our work to stimulate more theoretical and experimental studies to follow in this direction.
 
 \section{Conclusions}
\label{sec_conclusions}

In this work we have put forward a simple continuous biomathematical model allowing for the description of the formation of hypoxic hypercellular regions around necrotic cores in glioblastoma multiforme: the so-called pseudopalisades.  Pseudopalisading necrosis and microvascular hyperplasia are two of the most powerful predictors of poor prognosis among invasive gliomas. Rather than mere morphologic markers, these structures are almost certainly mechanistically instrumental in the accelerated growth properties that characterize the transition from a lower to a higher grade astrocytic glioma, such as glioblastoma. 

\par

Our biomathematical model quantifies the migration process influenced by the phenotypic switch of glioblastoma cells under hypoxic conditions caused by vaso-occlusive episodes. 
%Although a more detailed description should take into account the complex spatio-temporal dynamics of palisades and its surrounding neovasculature, 
In spite of the fact that tumor forms a three dimensional structure with a very complex dynamics, our numerical simulations clearly evidence that it is possible to observe the dependence between palisade lifetime, its necrotic core and characteristic width related to the distance from a thrombotic to a functional vessel within a minimal mathematical model. 

\par

We have also provided quantitative metrics of the characteristic times of formation and lifetimes of pseudopalisades after a vaso-occlusion event as well as their width and necrosis. We stress that these important metrics are very challenging to measure {\em in vivo} because of the small spatial scale of the phenomena involved and the fact that biopsies do not provide dynamical information in time but only a fixed image for a given time. In addition, our results indicate that smaller pseudopalisades would show shorter lifetimes and reduced necrotic cores while larger pseudopalisades would display longer lifetimes until reaching a threshold around the 8th day after the vaso-occlusion event. From our calculations, pseudopalisades tend to present a characteristic width of 200 $\mu$m when blood vessel distances are larger than 500 $\mu$m (within the range explored up to $700$ $\mu$m).

\par

Furthermore, our simulations show that palisading waves lead to a faster invasion than that by a pure random motion invasion with unlimited resources, implying that the mechanism of vessel break up might accelerate the glioblastoma progression. This suggests that targeting vaso occlusion events, probably by means of anticoagulants such as low molecular weight heparine, or other antithrombotic agents, might be of potential interest for therapeutics aimed at delaying glioblastoma growth. 

\par 
 
\begin{acknowledgements}
We wish to thank Marcial Garc\'{i}a Rojo and Cristina Murillo (Servicio de Anatom\'{\i}a Patol\'ogica, Hospital General Universitario de Ciudad Real, Spain) for fruitful discussions. We wish to thank D. Brat (Emory University, USA) for his support and for giving us the photographs used for Figs. \ref{fig:minipalisade_Brat} and \ref{fig:largepalisade_Brat}. 
This work has been supported by grants MTM2009-13832 (Ministerio de Ciencia e Innovaci\'on, Spain) and PEII11-0178 (Junta de Comunidades de Castilla-La Mancha, Spain).  
\end{acknowledgements}

\end{document}